\documentclass{appolb}
\usepackage{graphicx}
\usepackage{bm}
\usepackage[style=numeric,sorting=none]{biblatex} 
\begin{document}
\title{{\bf FCC-$\bm{\mu}\mathbf e$}: High Energy, High Luminosity Muon-Electron Collider \& Neutrino Factory at CERN%
\thanks{Contribution to the proceedings of  Workshop on Ultra-cold Muons for the LHC, Sep 24--26, 2026, at AGH University of Krak\'ow --   https://indico.cern.ch/event/1681458/}%
}
\author{Krzysztof PIOTRZKOWSKI
\address{AGH University of Krak\'ow, Faculty of Physics and Applied Computer Science, Al. Mickiewicza 30, 30--055 Kraków, Poland}
}

\maketitle
\begin{abstract}
The concept is presented of making at the FCC-$ee$ (or CEPC) facility $\mu^+e^-$ collisions at the centre-of-mass energy between 0.6 and 2 TeV and the luminosity of about  $10^{34}$cm$^{-2}$s$^{-1}$, concurrently to the $e^+e^-$ collisions. First, a concise scientific motivation is given, including the exciting prospect of measurements of the Higgs boson self-interaction, and is followed by a presentation of the FCC-$\mu e$ layout -- from the source of ultra-cold anti-muons, through the muon booster, to the main muon ring. Basic considerations are also given regarding the machine-detector interface and the performance of the FCC-$\mu e$ and associated neutrino factory. The paper concludes with an outlook underlying relevance of such anti-muon beams for any future high energy storage ring as well as for neutrino science.  Finally, the exciting case of $\mu^+e^-$ collisions at LEP3 is briefly discussed in the appendix.
\end{abstract}
  
\section{Introduction and motivation}

The concept of FCC-$\mu e$ is directly based on the ideas presented in the $\mu$TRISTAN proposal~\cite{Hamada:2022mua}, and has also profited from recent developments of LH$\mu$C~\cite{white-paper,Piotrzkowski:2026gjg}. The first considerations of the $\mu^+e^-$ collisions at FCC and CEPC should also be noted~\cite{Akturk:2024jbl}. In short, such a facility would provide high luminosity $\mu^+p$ collisions at the centre-of-mass energy $\sqrt{s_{\mu e}}$ between 0.6 and 2~TeV, opening very significant and complementary, with respect to the FCC-$ee$, research directions. Furthermore, as a free bonus, the FCC-$\mu e$ complex would produce very powerful neutrino beams, offering in Europe a potent and unique research platform for neutrino science.

There are three main reasons why the prospect of the FCC-$\mu e$ is very exciting. First and foremost, in contrast to FCC-$ee$ (due to its too small centre-of-mass energy), it offers direct access to 
the Higgs boson self-interactions via production of the Higgs boson pairs. Second, the single Higgs boson production via gauge boson fusion provides means for competitive and complementary measurements of the Higgs couplings\footnote{In a similar manner to the LH$e$C and LH$\mu$C cases, but yet with simpler final states, without the proton remnants.}~\cite{Hamada:2022mua,Lu:2020dkx}. Third, the initial state $\mu^+e^-$ offers unique sensitivities to charged lepton flavour violation (CLFV) processes~\cite{Lu:2020dkx} and interesting production mechanisms for supersymmetric particles~\cite{Hamada:2022mua}, for example.
In addition, strongly asymmetric beam energies are attractive for the search for long-lived particles~\cite{Hamada:2022mua}, and in contrast to the $e^+e^-$ and $\mu^+\mu^-$ colliders, at FCC-$\mu e$ the $s$-channel annihilation processes are absent, often resulting in a significant reduction of background levels.

The production of the Higgs boson pairs has been extensively studied for all future high energy lepton colliders, and in particular for the $\mu^+\mu^-$ case. For the main production mechanism proceeding via the $W$ boson fusion, a simple study was performed at the Monte Carlo generator level, assuming $H\rightarrow b\bar{b}$ decays. The "observed" LO parton-level cross-sections were calculated by requesting that all four $b$-quarks have pseudo-rapidities within the range $|{\eta_b}|< 5$, and that two $b\bar{b}$ invariant masses are consistent with the Higgs boson mass. The
cross-sections obtained for the FCC-$\mu e$ signal and irreducible backgrounds\footnote{At FCC-$\mu e$ that involves $\mu^+ e^- \rightarrow \bar{\nu}_\mu \nu_e b\bar{b}b\bar{b}$ and $\mu^+ e^- \rightarrow \bar{\nu}_\mu \nu_e Hb\bar{b}$ followed by the $H\rightarrow b\bar{b}$ decay. The corresponding final states are also relevant for the $\mu^+\mu^-$ case.} are very close to the ones of the $\mu^+\mu^-$ collider  at the corresponding centre-of-mass energy~\cite{hamzeh3}. Therefore, by re-scaling the limits obtained for the $\mu^+\mu^-$ collider~\cite{Andreetto:2024rra}, one can very tentatively obtain the FCC-$\mu e$ sensitivities to the Higgs boson trilinear self-coupling $\lambda_3$: for $\sqrt{s_{\mu e}}=1$~TeV, assuming an integrated luminosity of 1~ab$^{-1}$, one then may expect a precision close to that at the HL-LHC~\cite{ATLAS:2025eii}, which however would significantly improve for the $\mu^+$ beam polarisation $P_{\mu^+}=80\%$, as the cross-sections are proportional to $1+P_{\mu^+}$; for $\sqrt{s_{\mu e}}\simeq2$~TeV, assuming an integrated luminosity of 3~ab$^{-1}$, one then may expect ("symmetrised") precision of about 20\% for non-polarised anti-muons, which in case of $P_{\mu^+}=80\%$ would still significantly improve!
It should be noted that the $s$-channel $ZH$ background is absent at FCC-$\mu e$. Finally, the Higgs boson pairs will be produced also via the $ZZ$ fusion, $\mu^+ e^- \rightarrow \mu^+ e^- HH$, and its cross-section is only about 13\% of the $WW$ one, but possible kinematical constraints might help to further reduce backgrounds and reach a good sensitivity.
\section{FCC-$\mu e$ concept}
\begin{figure}[h]
\centering
\includegraphics[width=0.96\textwidth]{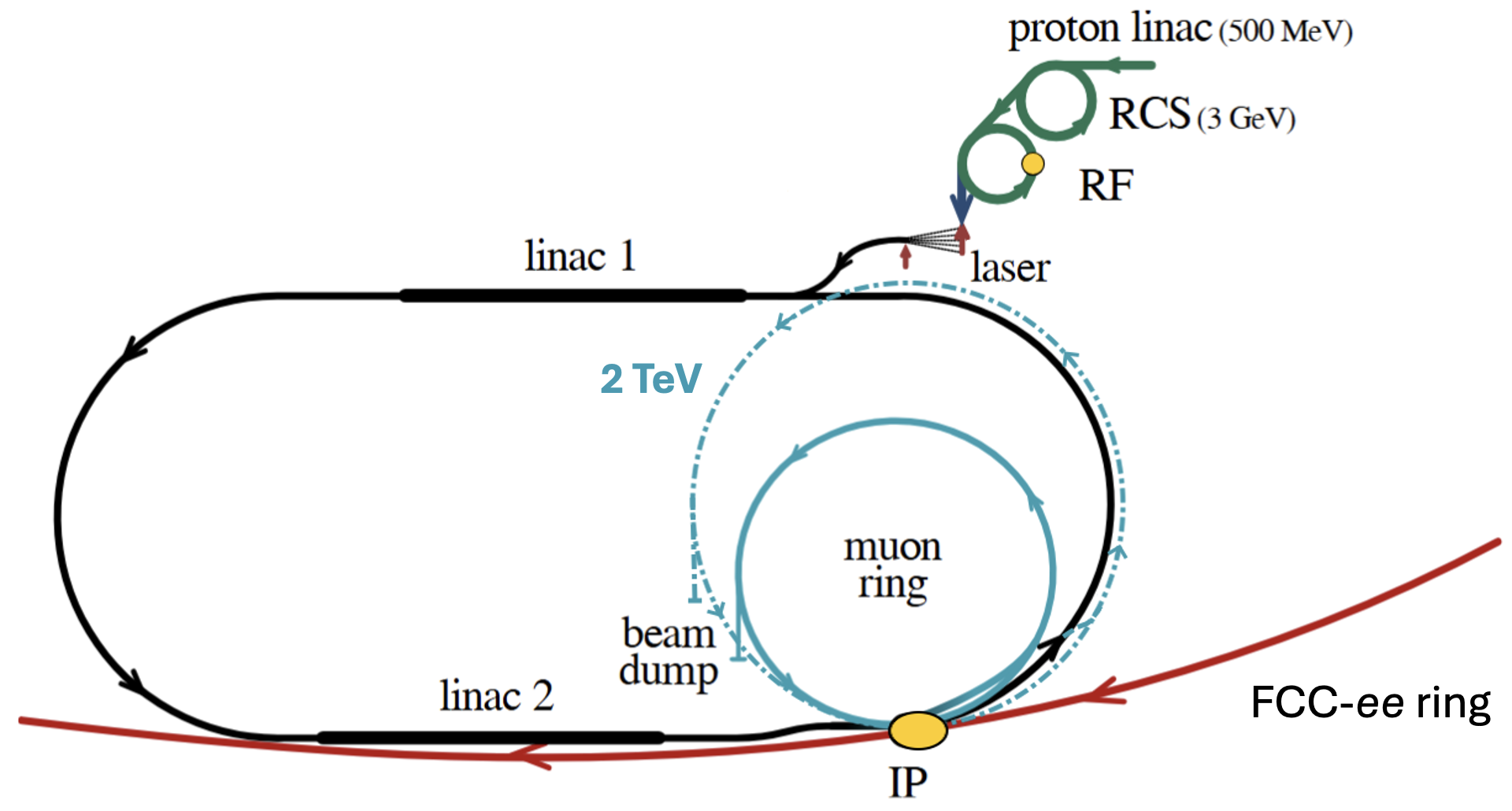}
\caption{Layout of the FCC-$\mu e$ concept, based on the $\mu$TRISTAN proposal~\cite{Hamada:2022mua}. The $\mu^+$ booster is depicted in black, while the 1~TeV muon storage ring is shown in light blue and the dash-dotted circle represents the 2 TeV muon ring.
\label{fig:i}}
\end{figure}

In the following, it is assumed that a 2~TeV anti-muon beam collides head-on with the electron beam of the FCC-$ee$, at the PL interaction point (IP), see Fig.~\ref{fig:i}. As in the LH$\mu$C concept~\cite{Piotrzkowski:2026gjg}, the $\mu^+$ booster containing two linacs, inherited from LH$e$C, has a circumference of approximately 9~km, while the 2 TeV muon storage ring's is of about 6~km. In addition, also a 5~TeV beam is discussed for the last stage of the FCC-$\mu e$ operation, assuming the use of the HTS magnets.\\

\vspace{-3mm}
\label{layout}
\vspace{2mm}

In general, the $\mu^+e^-$ collisions could be made concurrently with the nominal FCC-$ee$ running because the extra beam-beam effects can be kept at acceptable levels -- moreover,  "stand-alone" running of FCC-$\mu e$, which might be of great value, is also considered. 
The ultra-cold $\mu^+$ source in J-PARC has achieved outstanding normalised transverse emission, $\epsilon_n(\mu)$, of approximately 0.5~$\mu$m~\cite{Aritome:2024rlu}. If such an emittance is maintained for the required high-intensity source, and through all beam transport and acceleration, then $\epsilon(\mu)^2<\epsilon_x\epsilon_y$ and the lateral beam sizes of the electron beam at IP(PL), $\sigma_x=\sqrt{\beta^*_x\epsilon_x}$ and $\sigma_y=\sqrt{\beta^*_y\epsilon_y}$, can be matched by the $\mu^+$ beam sizes. Finally, the anti-muon bunch length is expected to be shorter than the electron one.
\section{$\mu^+p$ interaction region}
The $\mu^+e^-$ interaction region is very different from the $e^+e^-$ ones:
\begin{itemize}
    \item There is no "immediate" beam-pipe splitting in the central region as anti-muons collide head-on with electrons -- the $\mu^+$ beam gets separated at a larger distance from the IP;
    \item the central beam pipe is larger, with about 2 cm radius\footnote{This should not much affect the flavour tagging as the FCC-$\mu e$ centre-of-mass energy is at least 2 times bigger than that of the FCC-$ee$, hence correspondingly bigger vertex displacements, on average.}, and significantly longer as first quadrupoles are further away from the IP (see below);
    \item the non-colliding, "parallel" positron beam must be passed through, at a lateral distance of about 1 cm;
    \item synchrotron radiation might be somewhat weaker due to less violent focussing, but the $\mu^+$ side requires installing a "nozzle" -- special shielding (mostly for the central detector) against the flux of high energy positrons due to the upstream beam muon decays -- in addition, more beam-line shielding will also be needed; one should note however that the Muon Collider Collaboration has studied in-depth the necessary solutions both for the 2 and 5~TeV case.
\end{itemize}
\section{FCC-$\mu e$ performance and running}
Since the horizontal and vertical $\mu^+$ emittances are similar and collisions are head-on, the colliding beams need to be less flat at the IP, i.e. the ratio of lateral beam sizes, $\sigma_x/\sigma_y$, should be smaller by an orders of magnitude than for the $e^+e^-$ case. Using a 4~MW proton driver, one obtains excellent FCC-$\mu e$ performance, see Tab.~\ref{tab:FCC-ue}. It should be noted that $\mu^+e^-$ luminosity at LEP3 is expected to be similar or even higher -- see Appendix.  The $\mu^+$ train stored in the main ring in this case contains 800 (20) bunches separated by 0.025 (1)~$\mu$s for running with the 45.6 (120)~GeV electron beam, and the repetition rate is 12.5~Hz. The beam power at the main ring is 6.4~MW. This requires the $\mu^+$ yield at a level of 20 and 4\% of the "ultimate" one in the $\mu$TRISTAN proposal for the lower and higher electron beam energies, respectively.
\begin{table}[h]
\centering
\caption{Example of FCC-$\mu e$ parameters for 2~TeV anti-muons on 45.6 (120)~GeV electrons with the assumed emittances $\epsilon_x = 0.7~ (2.1)$~nm and $\epsilon_y= 1.4~(2.1)$~pm.}
\vspace{2mm}
\label{tab:FCC-ue}
\begin{tabular}{lcccc}
\hline
\textbf{FCC-$\mu e$ parameters} & \multicolumn{4}{c}{\textbf{Value}} \\
\hline
CoM energy, $\sqrt{s_{\mu e}}$ [TeV] & \multicolumn{2}{c}{0.6} & \multicolumn{2}{c}{0.98} \\
Luminosity [$10^{34}$ cm$^{-2}$s$^{-1}$] & \multicolumn{2}{c}{4.5} & \multicolumn{2}{c}{0.63} \\
Collision rate [MHz]&\multicolumn{2}{c}{40}&\multicolumn{2}{c}{1} \\
Beam sizes at IP, $\sigma_{x}~/~\sigma_{y}$ [$\mu$m] &\multicolumn{4}{c}{13 / 0.22}  \\
\hline
\textbf{Beam parameters} & $e^-$ & $\mu^+$ & $e^-$ & $\mu^+$\\
\hline
$\beta^*_x$~[m] / $\beta^*_y$ [mm] & $0.24\,/\,3.5$ & $6.5\,/\,2$ & $0.08\,/\,2.3$ & $6.5\,/\,2$ \\
Bunch population [$10^{10}$]& 20 & 0.2 & 22 & 1\\
Beam-beam parm $\xi_h\,/\,\xi_v\,$[$10^{-2}$]& $0.5\,/\,12$ & $85\,/\,1$ & 1 /$\,15$ & $94\,/\,2$\\
\hline
\vspace{0mm}
\end{tabular}
\end{table}
Running with the 182.5~GeV electron beam, despite offering the highest centre-of-mass energy, is of secondary importance because of the relatively small feasible luminosity of about $2.10^{33}$~cm$^{-2}$s$^{-1}$. Note that in general the expected luminosities for the 5~TeV $\mu^+$ beam are significantly higher than for the 2~TeV case due to the smaller $\mu^+$ beam emittance and increased bunch population (the repetition rate is then only 5~Hz). 
%

The FCC-$\mu e$ luminosity should be kept constant ("levelled") in a similar way as it is done at the LHC -- while the $\mu^+$ beam intensity drops, the horizontal beam sizes are correspondingly shrunken by stronger focussing. 
Once the technical feasibility of such a fast levelling (within 80~ms) is confirmed, it would be very interesting to investigate extending it further to allow even lower repetition rates, hence increasing bunch populations.
It is very likely that the FCC-$\mu e$ luminosity can still be significantly boosted for the case of "stand-alone" running, with no positron beam and no collisions in the $e^+e^-$ experiments\footnote{In this case the beam-beam limits might become higher for the IP(PL) and hence the $\mu^+$ bunch populations increased.}. 
The concurrent running of FCC-$\mu e$ using a 2~TeV beam should result in a total integrated luminosity (including all electron beam energies) of several ab$^{-1}$. In case of running with a 5~TeV $\mu^+$ beam, the expected total integrated luminosity might still be significantly higher, especially when the "stand-alone" FCC-$\mu e$ operations are also possible once the proper FCC-$ee$ programme ends.

Note that the 5~TeV stage of FCC-$\mu e$ would also serve as an ultimate demonstration of the HTS technology for the FCC-$hh$ project, as well as would "for free" ensure availability of high luminosity 32 TeV $\mu^+p$ collisions\footnote{A 2~TeV anti-muon beam then provides collisions at $\sqrt{s_{\mu p}}=20$~TeV.} resulting in a very significant extension of the scientific scope of FCC-$hh$.
\section{Neutrino factory}
Last but not least, from very beginning the FCC-$\mu e$ complex will produce "for free" $\nu_e$ and $\bar{\nu}_\mu$ beams -- very intense and highly collimated in three different directions -- along two booster linacs as well as the straight section of the main ring. The energies of the decaying anti-muons are shared between complementary bands, from 1~GeV to the top energy. It is worth noting that the energy $E_{\mu^+}$ of the anti-muons decayed in the booster is distributed, to a very good approximation, as $1/E_{\mu^+}$.

Assuming the above parameters of the 2~TeV $\mu^+$ complex and that it will run throughout all year, the annual numbers of the anti-muon decays are as follows for the setup with the 45.6~GeV electron beam: above $2.10^{20}$ for the two linacs combined and about $3.10^{19}$ for the main ring, assuming a 300~m long (PL) straight section. For the 5~TeV case, not only the neutrinos' energy reach will be significantly higher, but possibly also their fluxes. 
A striking time structure of the produced neutrinos will allow one to easily associate any detected neutrino with a proper energy-band of the parent muon. Furthermore, in the design phase, there should be quite a lot of flexibility in the choice of booster beam directions. All in all, such a facility should be considered a neutrino factory.
\section{Outlook}
Adding high energy, high luminosity $\mu^+e^-$ collisions to the FCC-$ee$ (or CEPC) facility would provide a versatile and profound extension of its scientific programme, in addition to offering "for free" a unique and powerful platform at CERN for neutrino science.
Furthermore, in the long term, the FCC-$\mu e$ project would open fascinating opportunities to extend the scientific scope of the future FCC-$hh$ facility by adding a 32~TeV $\mu^+p$ collider at a fractional cost.

 Such $\mu^+$ beams would be of huge added value for any future high energy storage ring, be it the HL-LHC, FCC-$ee$, LEP3, HE-LHC, or FCC-$hh$, in each case with an accompanying bonus neutrino factory. In the appendix, the very interesting LEP3 case is briefly discussed.


\section*{Acknowledgements}
{
\noindent
The author thanks Hamzeh Khanpour for sharing the results of his cross-section calculations with {\sc MadGraph}.

K.P. is supported by the NCN grant 2023/49/B/ST2/03273 and the NAWA grant BPN/PPO/2021/1/00011.}

\printbibliography

@misc{white-paper,
  author       = { {HELP4CERN Study Group}}, 
  title        = "{White Paper on High Energy, High Luminosity Lepton-Hadron Colliders at CERN}",
  note         = {In preparation},
  year         = {2026}
}

@misc{hamzeh3,
  author       = { {H. Khanpour}}, 
  title        = "{Private communication}"
}

@article{Anastopoulos:2025jyh,
    author = "Anastopoulos, C. and others",
    title = "{LEP3: a high-luminosity $e^+e^-$ Higgs and electroweak factory in the LHC tunnel}",
    eprint = "2504.00541",
    archivePrefix = "arXiv",
    primaryClass = "physics.acc-ph",
    doi = "10.1088/1361-6471/ae5728",
    journal = "J. Phys. G",
    volume = "53",
    number = "4",
    pages = "040501",
    year = "2026"
}

@article{Akturk:2024jbl,
    author = "Akturk, D. and Dagli, B. and Sultansoy, S.",
    title = "{Muon ring and FCC-ee/CEPC based antimuon-electron colliders}",
    eprint = "2403.17034",
    archivePrefix = "arXiv",
    primaryClass = "physics.acc-ph",
    doi = "10.1209/0295-5075/ad5e1a",
    journal = "EPL",
    volume = "147",
    number = "2",
    pages = "24001",
    year = "2024"
}

@article{ATLAS:2025eii,
    author = "Aad, Georges and others",
    collaboration = "ATLAS, CMS",
    title = "{Highlights of the HL-LHC physics projections by ATLAS and CMS}",
    eprint = "2504.00672",
    archivePrefix = "arXiv",
    primaryClass = "hep-ex",
    reportNumber = "ATL-PHYS-PUB-2025-018, CMS-HIG-25-002",
    month = "4",
    year = "2025"
}

@article{Lu:2020dkx,
    author = "Lu, M. and Levin, Andrew Michael and Li, Congqiao and Agapitos, Antonios and Li, Qiang and Meng, Fanqiang and Qian, Sitian and Xiao, Jie and Yang, Tianyi",
    title = "{The physics case for an electron-muon collider}",
    eprint = "2010.15144",
    archivePrefix = "arXiv",
    primaryClass = "hep-ph",
    doi = "10.1155/2021/6693618",
    journal = "Adv. High Energy Phys.",
    volume = "2021",
    pages = "6693618",
    year = "2021"
}

@inproceedings{Piotrzkowski:2026gjg,
    author = "Piotrzkowski, K.",
    collaboration = "HELP4CERN Study Group",
    title = "{LH$\mu$C: A high energy, high luminosity muon-hadron collider at CERN (on behalf of the HELP4CERN Study Group)}",
    eprint = "2610.35154",
    archivePrefix = "arXiv",
    primaryClass = "hep-ex",
    month = "10",
    year = "2026"
}

@article{Andreetto:2024rra,
    author = "Andreetto, Paolo and others",
    title = "{Aspects of Higgs Physics at a $\sqrt{s}=3$ TeV Muon Collider with detailed detector simulation}",
    eprint = "2405.19314",
    archivePrefix = "arXiv",
    primaryClass = "hep-ex",
    reportNumber = "FERMILAB-CONF-24-0679-PPD",
    doi = "10.1140/epjc/s10052-025-13923-6",
    journal = "Eur. Phys. J. C",
    volume = "85",
    number = "3",
    pages = "221",
    year = "2025"
}

@article{Ahmadova:2025vzd,
    author = "Ahmadova, F. and others",
    title = "{The Large Hadron electron Collider as a bridge project for CERN}",
    eprint = "2503.17727",
    archivePrefix = "arXiv",
    primaryClass = "hep-ex",
    month = "3",
    year = "2025",
    note = "Accepted for publication in $J.$ $Phys.~G$"
}

@article{Aritome:2024rlu,
    author = "Aritome, S. and others",
    title = "{Acceleration of Positive Muons by a Radio-Frequency Cavity}",
    eprint = "2410.11367",
    archivePrefix = "arXiv",
    primaryClass = "physics.acc-ph",
    doi = "10.1103/PhysRevLett.134.245001",
    journal = "Phys. Rev. Lett.",
    volume = "134",
    number = "24",
    pages = "245001",
    year = "2025"
}

@article{Hamada:2022mua,
    author = "Hamada, Y. and Kitano, Ryuichiro and Matsudo, Ryutaro and Takaura, Hiromasa and Yoshida, Mitsuhiro",
    title = "{$\mu$TRISTAN}",
    eprint = "2201.06664",
    archivePrefix = "arXiv",
    primaryClass = "hep-ph",
    reportNumber = "KEK-TH-2385",
    doi = "10.1093/ptep/ptac059",
    journal = "PTEP",
    volume = "2022",
    number = "5",
    pages = "053B02",
    year = "2022"
}

\vspace{7mm}

\section*{Appendix}
LEP3 is the proposed "high-luminosity $e^+e^-$  Higgs \& electroweak factory in the LHC tunnel", in case the FCC-$ee$ is not to be built~\cite{Anastopoulos:2025jyh}. A similar layout to the LH$\mu$C one is considered at LEP3 to deliver $\mu^+e^-$  collisions (simultaneously with the $e^+e^-$ ones in this case), where the proton ring is replaced by the electron one. At LEP3-$\mu e$, as the 182.5~GeV beam is not possible, the $\sqrt{s_{\mu e}}$ range  will be from 0.6 to 0.96~TeV and from 0.95 to 1.5~TeV, for the 2 and 5~TeV  $\mu^+$ beams, respectively. 

As shown in Tab.~\ref{tab:LEP3-ue}, the expected performance of LEP3-$\mu e$ is excellent -- the luminosity at 0.6~TeV is significantly higher than
for the FCC-$\mu e$ case! The $\mu^+$ bunch trains stored in the main ring contain 200 (5) bunches separated by 0.1 (4)~$\mu$s to collide with the 45.6 (115)~GeV electron beam, and the repetition rate is also 12.5~Hz. This requires the $\mu^+$ yield at a level of 25 and 5\% of the "ultimate" one in the $\mu$TRISTAN proposal for the lower and higher electron beam energies, respectively. This means that during the concurrent running the
"intrinsic" neutrino factory might deliver slightly stronger neutrino beams than in the FCC-$ee$ case.\\
\begin{table}[h]
\centering
\caption{Two examples of $\mu^+e^-$ collisions at LEP3 of 2~TeV anti-muons with 45.6 and 120~GeV electrons with the beam parameters according to Ref.~\cite{Anastopoulos:2025jyh}.
}
\vspace{2mm}
\label{tab:LEP3-ue}
\begin{tabular}{lcccc}
\hline
\textbf{LEP3-$\mu e$ parameters} & \multicolumn{4}{c}{\textbf{Value}} \\
\hline
CoM energy, $\sqrt{s_{\mu e}}$ [TeV] & \multicolumn{2}{c}{0.6} & \multicolumn{2}{c}{0.96} \\
Luminosity [$10^{34}$ cm$^{-2}$s$^{-1}$] & \multicolumn{2}{c}{7.2} & \multicolumn{2}{c}{0.52} \\
Collision rate [MHz]&\multicolumn{2}{c}{10}&\multicolumn{2}{c}{0.25} \\
Beam sizes at IP, $\sigma_{x}~/~\sigma_{y}$ [$\mu$m] &\multicolumn{4}{c}{13 / 0.22}  \\
\hline
\textbf{Beam parameters} & $e^-$ & $\mu^+$ & $e^-$ & $\mu^+$\\
\hline
$\beta^*_x$~[m] / $\beta^*_y$ [mm] & $0.24\,/\,3.5$ & $6.5\,/\,2$ & $0.08\,/\,2.3$ & $6.5\,/\,2$ \\
Bunch population [$10^{10}$]& 26 & 1 & 25 & 3\\
Beam-beam parm $\xi_h\,/\,\xi_v\,$[$10^{-2}$]& $8\,/\,13$ & $110\,/\,2$ & 2 /$\,13$ & $108\,/\,2$\\
\hline
\vspace{1mm}
\end{tabular}
\end{table}

These preliminary observations imply great scientific potential for LEP3-$\mu e$, which is  complementary to the nominal LEP3 research programme, and which will also include a powerful platform at CERN for neutrino science.
That gives yet another reason, in case FCC-$ee$ cannot be realised, to build first LH$e$C~\cite{Ahmadova:2025vzd} and then LEP3.

\end{document}